\documentstyle[psfig,epsfig,sprocl]{article}

\def\lsim{\:\raisebox{-0.5ex}{$\stackrel{\textstyle<}{\sim}$}\:}

\def\21{$SU(2) \otimes U(1) $}

\def\ib#1#2#3{           {\it ibid. }{\bf #1} (19#2) #3}

\def\nps#1#2#3{        {\it Nucl. Phys. B (Proc. Suppl.) }{\bf #1} (19#2) #3} 
\def\np#1#2#3{           {\it Nucl. Phys. }{\bf #1} (19#2) #3}
\def\pl#1#2#3{           {\it Phys. Lett. }{\bf #1} (19#2) #3}
\def\pr#1#2#3{           {\it Phys. Rev. }{\bf #1} (19#2) #3}
\def\prep#1#2#3{         {\it Phys. Rep. }{\bf #1} (19#2) #3}
\def\prl#1#2#3{          {\it Phys. Rev. Lett. }{\bf #1} (19#2) #3}

\def\n.c.#1#2#3{         {\it Nuovo Cim. }{\bf #1} (19#2) #3}
\def\r.n.c.#1#2#3{       {\it Riv. del Nuovo Cim. }{\bf #1} (19#2) #3}
\def\sjnp#1#2#3{         {\it Sov. J. Nucl. Phys. }{\bf #1} (19#2) #3}

\def\ppnp#1#2#3{           {\it Prog. Part. Nucl. Phys. }{\bf #1} (19#2) #3}

\def\ip{in preparation}
\def\ne{\hbox{$\nu_e$ }}
\def\nm{\hbox{$\nu_\mu$ }}
\def\nt{\hbox{$\nu_\tau$ }}

\def\mnt{\hbox{$m_{\nu_\tau}$ }}
\def\fig#1{{Fig. (\ref{#1})}}
\def\beq{\begin{equation}}
\def\eeq{\end{equation}}
\def\bef{\begin{figure}}
\def\eef{\end{figure}}
\def\eq#1{{eq. (\ref{#1})}}
\begin{document}

\title{ Supersymmetry without R-Parity
\footnote{Review talk given at the International Workshop on Quantum
Effects in the MSSM, held at Barcelona, Sep. 1997.}}
\author{ Jos\'e W. F. Valle 
\footnote{E-mail valle@flamenco.ific.uv.es}}
\address{ Instituto de F\'{\i}sica Corpuscular - C.S.I.C.\\
Departament de F\'{\i}sica Te\`orica, Universitat de Val\`encia\\
46100 Burjassot, Val\`encia, SPAIN}
\maketitle
\abstracts{
I discuss the motivations for supersymmetry, focussing on models with
broken R--parity and lepton number. After describing the main
theoretical features of these models, I discuss some of the signals
expected at colliders such as Tevatron, LEP II, NLC and LHC.  }

\section{Introduction}

The Standard Model (SM) is very successful in describing the
fundamental elementary particle interactions, except possibly
neutrinos. It leaves many unanswered questions and theoretical
problems.  A basic ingredient in the SM is the breaking of the
electroweak symmetry via the Higgs mechanism. One of its most
outstanding puzzles is the fact that the mass of the SM Higgs boson is
unstable against quantum corrections, a fact known as the hierarchy
problem.  One of the main theoretical motivations for supersymmetry
(SUSY) is that it allows for a stable hierarchy between the
electroweak scale $m_{weak}$ responsible for the W and Z masses and
the mass scale of unification.  If SUSY holds as a symmetry down to
the scale $ M_{SUSY} \sim m_{weak}$, then the Higgs mass is stabilized
under radiative corrections. This happens because the loops containing
standard particles are partially cancelled by those containing
supersymmetric particles. Supersymmetry now appears as the most
natural and well-founded solution to the hierarchy problem, at least
in a technical sense.

Another drawback of the SM is that the weak, electro-magnetic and
strong interactions are characterized by couplings of different
strength. Supersymmetry also allows in an elegant and natural way the
unification of the three gauge couplings, when evolved via the
renormalization group equations from $m_{weak}$ up to the unification
scale \cite{GUT}.

The minimal realization of SUSY is the so-called Minimal
Supersymmetric Standard Model (MSSM) \cite{mssm}. Although it has the
advantage of being the simplest, the MSSM is an {\sl ad hoc} choice,
which is by no means mandatory. I discuss the simplest effective way
to include R--parity violation which mimics the main features of more
complete dynamical models where this violation happens due to the
non-zero expectation values of sneutrinos. I mention some of the
physics motivations and potential of various extensions of the MSSM
with broken R--parity. Since neutrinos typically have mass in these
models, many of the related phenomena are deeply related to the
physics of weak interactions and the properties of neutrinos.

\section{ Supersymmetry and the MSSM}

Supersymmetry has several attractive theoretical features.  For
example, as already mentioned, it allows the resolution of the
hierarchy problem, if it survives down to the weak scale, leading to
the possibility of discovering a whole plethora of SUSY particles with
masses in the TeV scale at the new generation of colliders LEP II and
LHC. Moreover, SUSY allows for a very elegant way to break the
electroweak symmetry via radiative corrections  \cite{mssmrad}.

\begin{figure}[t]
\centerline{\protect\hbox{\psfig{file=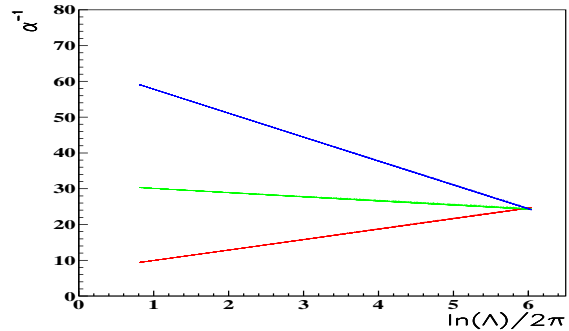,height=5cm,width=8.5cm}}}
\vglue -0.3cm
\caption{Gauge coupling unification with the latest experimental data.}
\label{gunif}
\end{figure}

Another attractive feature of SUSY is that precision measurements of
the three gauge couplings performed at the CERN $e^+e^-$ collider LEP
as well as neutral current data \cite{PDG96} are all in good agreement
with the MSSM--GUT with the SUSY scale $M_{SUSY}\lsim 1$ TeV
\cite{gaugUnifRecent}. This is illustrated in \fig{gunif}
taken from ref. \cite{hunif2}.
It is important to stress that the unification scale in SUSY--GUT is
high enough to predict a proton decay rate slower than present
experimental limits, as opposed to the non--SUSY GUTs, where the
proton decays too fast.

The simplest SUSY model is the MSSM \cite{mssm}. This model realizes
SUSY in the presence of a discrete R--parity symmetry. Under this
symmetry all standard model particles are even while their partners
are odd. As a result of this selection rule SUSY particles are only
produced in pairs, with the lightest of them being stable. In the MSSM
the Lightest SUSY Particle (LSP for short) is typically a neutralino,
for most choices of SUSY parameters.  It has been suggested as a
candidate for the cold dark matter of the universe and several methods
of detection at underground installations have been suggested
\cite{chicdm}.

However, one should not forget that R--parity is postulated {\sl ad
hoc}, without a deep theoretical basis. Moreover there are other ways
to explain the cold dark matter via the axion. Last, but not least,
hot dark matter is needed in any case, not to to mention other
existing puzzles in neutrino physics, such as the solar neutrino deficit,
which require non-zero neutrino mass. From this point of view the
emphasis of the simplest MSSM picture would seem exaggerated.

\section{Supersymmetry with Explicitly Broken R--Parity}

R--parity could well be broken via {\sl tri-linear} superpotential
couplings of the type
\beq
\lambda E L L, \:\:\: \lambda' D Q L, \: \: \:  \mbox{and} \:\:\:  
\lambda'' U D D\,,
\eeq
where the U, D, and E are SU(2) singlet superfields corresponding to
the right-handed u, d quarks as well as charged leptons. These could
arise from gravitational effects, in which case they are expected to
be tiny \cite{rpgrav}.

\begin{figure}[t]
\centerline{\protect\hbox{
\psfig{file=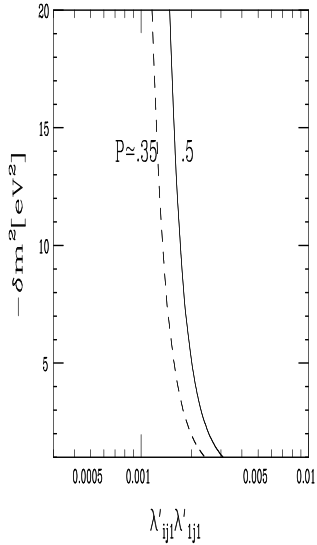,height=5cm,width=8.5cm,angle=90}}}
\vglue -.7cm
\caption{SN1987A bounds on tri-linear R--parity violation. }
\label{fcncprob2}
\end{figure}

There are strong constraints on many of the corresponding couplings,
especially the baryon-number violating ones, because of proton
stability. There are also many bounds that follow from high energy
physics as well as nuclear physics experiments, such as nuclear double
beta decays \cite{heid}.  For a recent compilation see ref.
\cite{3lbounds}. There are, in addition some cosmological and
astrophysical limits. For example, preserving a cosmological baryon
asymmetry generated at the unification scale severely restricts
certain combinations of $\lambda's$, barring the existence of special
symmetries. Alternatively, the baryon asymmetry may also be created at
the weak scale. As for astrophysical limits I mention a limit recently
derived in ref. \cite{rsusysn}. It is based on the observation that
the tri-linear couplings lead to flavour changing neutral current
neutrino interactions which may induce resonant massless-neutrino
conversions in a dense supernova medium. As shown in \fig{fcncprob2}
the restrictions that follow from the observed $\bar\nu_e$ energy
spectra from SN1987A are much more stringent than those obtained from
the laboratory. For the opposite sign of the neutrino mass square
difference $\delta m^2$ supernova $r$-process nucleosynthesis gives
complementary restrictions \cite{rsusysn}. Altogether, these disfavour
a leptoquark interpretation of the recent HERA anomaly \cite{HERA}.

A simpler and more interesting way to break is via {\sl bi-linear}
superpotential couplings.  In this case the superpotential is given by
\cite{epsrad,epsi,RPothers}
\begin{eqnarray} 
&&W=
 h_t \widehat Q_3 \widehat U_3\widehat H_u
+h_b \widehat Q_3 \widehat D_3\widehat H_d
+h_{\tau}\widehat L_3 \widehat R_3\widehat H_d \nonumber\\
&&\qquad\qquad\,+\mu\widehat H_u \widehat H_d
+\epsilon_3\widehat L_3 \widehat H_u 
\label{eq:Wsuppot}
\end{eqnarray}
where the first four terms correspond to the MSSM and the last one is
the bi-linear term which violates R--Parity and lepton number
explicitly (for three generations there would be three $\epsilon_i$).
This superpotential is motivated by models of spontaneous breaking of
R--Parity \cite{MASIpot3} \cite{RIV} \cite{sbrpothers} \cite{RPCHI}
\cite{RPLR}.  Contrary to a popular misconception, the bi-linear
violation of R--parity implied by the parameter $\epsilon_3$ is
physical and can not be rotated away \cite{DJV}.  Whichever way one
chooses to parametrize the model there is R--parity violation which
also implies a non-zero sneutrino vacuum expectation value $v_3$.

\begin{figure}[t]
\centerline{\protect\hbox{
\psfig{file=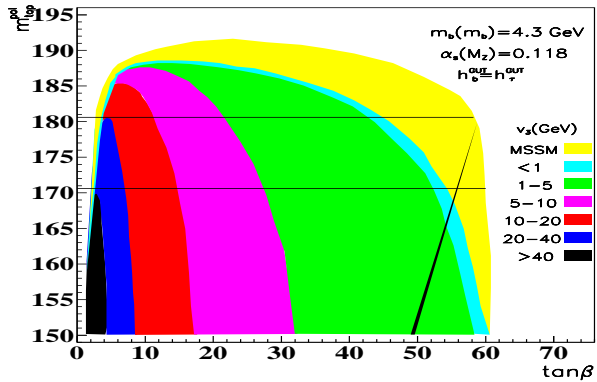,height=5.5truecm,width=8.5truecm}}}
\caption{Top quark mass versus $\tan\beta$ for different
values of the sneutrino VEV $v_3$.}
\label{aretop}
\end{figure}

One attractive feature of this model is that it allows the radiative
breaking of the electroweak symmetry with the simplest assumption of
universal soft SUSY breaking terms at unification \cite{epsrad}. In
contrast to the MSSM \cite{YukUnif}, however, this model allows for
the unification of the bottom and tau Yukawa couplings at the scale
$M_{GUT}$ where the gauge couplings unify \cite{hunif} for any value
of $\tan\beta$ provided $v_3$ is chosen appropriately \cite{hunif}.
This is illustrated in \fig{aretop}, taken from ref. \cite{hunif}.
In \fig{aretop} the bottom quark and tau lepton Yukawa couplings are
unified at $M_{GUT}$ and the horizontal lines correspond to the
$1\sigma$ experimental $m_t$ determination (for simplicity
$M_{SUSY}=m_t$ was assumed). The diagonal band at high $\tan\beta$
values corresponds to $t-b-\tau$ unification, expected in SO(10)
models.

Note that $\epsilon_3$ and the $v_3$ are related by a minimization
condition. As a result, if we adopt universal conditions for the soft
breaking parameters this model contains effectively a single extra
free parameter in addition to those of the minimal supergravity model.
In this case R--parity violation is induced radiatively, due to
the effect of the non-zero bottom quark Yukawa coupling $h_b$ in the
running of the renormalization group equations from the unification
scale down to the weak scale \cite{DJV}.

\begin{figure}[t]
\centerline{\protect\hbox{\psfig{file=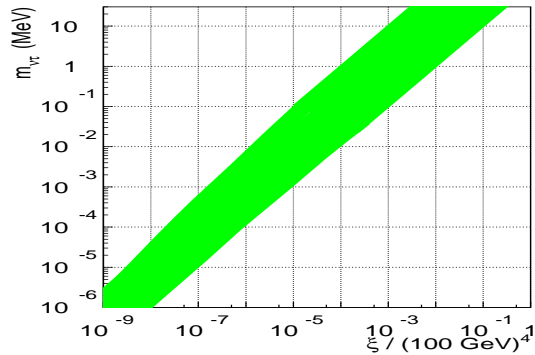,height=5cm,width=8cm}}}
\caption{Tau neutrino mass versus the effective parameter $\xi$ }
\vglue -0.3cm
\label{mnt_xi_ev}
\end{figure}

Another important property of the bi-linear model of R-parity breaking
is that it provides a very elegant mechanism for the origin of
neutrino mass which combines the virtues of the seesaw \cite{GRS} and
the radiative mechanisms \cite{zee.Babu88} of neutrino mass generation
\cite{fae}. The tau neutrino $\nu_{\tau}$ acquires a mass, due to the 
mixing between neutrinos and neutralinos If we stick to the simplest
unified supergravity version of the model with bi-linear breaking of
R--parity and universal boundary conditions for the soft breaking
parameters \cite{epsrad,RPothers}, then the {\sl effective} neutralino
mixing parameter $\xi \equiv (\epsilon_3 v_1 + \mu v_3)^2$
characterizing the violation of R--parity, either through $v_3$ or
$\epsilon_3$ will be small since contributions arising from {\sl
gaugino} mixing will cancel, to a large extent, those from {\sl
Higgsino} mixing.  This cancellation will happen automatically if the
soft breaking parameters are universal \cite{epsrad,DJV}. In this case
\mnt will be naturally small and radiatively calculable in terms of
the bottom Yukawa coupling $h_b$. This will explain the smallness of
the neutrino mass in this model.  The above scenario is a {\sl hybrid} of
the see-saw and radiative schemes of neutrino mass generation.  The r\^ole
of the right-handed mass is played by the neutralinos mass (which lies
at the weak scale) while the r\^ole of the Dirac mass is played by the
{\sl effective} neutralino mixing $\xi$ which is induced
radiatively. The \nt mass induced this way is directly correlated with
the magnitude of the effective parameter $\xi$.  In \fig{mnt_xi_ev} we
display the allowed values of $m_{\nu_{\tau}}$.

\begin{figure}[t]
\centerline{\protect\hbox{\psfig{file=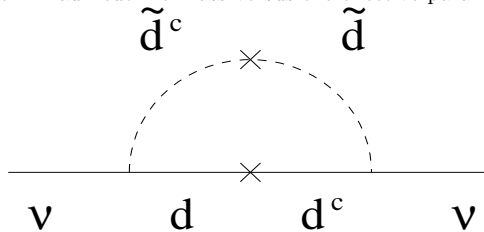,height=3.5cm,width=6.5cm}}}
\vglue -0.5cm
\caption{Scalar loop contributions to neutrino masses.  }
\vglue -0.3cm
\label{mnrad}
\end{figure}

It is important to notice that this happens for relatively large
values of the relevant model R--parity violation parameter
$\epsilon$. As a result many of the corresponding R-parity violating
effects can be sizeable even when \mnt is small \footnote{However,
\mnt can be as large as the present laboratory bound \cite{ntbound} 
in models with spontaneous breaking of R--parity.}.  Moreover there
can be striking effects of R--parity violation which do not require it
to have a large strength.  The obvious example is the fact that the
lightest neutralino decay will typically decay inside the detector,
unless the violation is really tiny as in \cite{rpgrav}.
Notice that \ne and \nm remain massless in this approximation. They
get masses either from scalar loop contributions 
\footnote{Here we use a different basis in which the bi-linear term
is removed but re-introduces a tri-linear term $D Q L$ whose
coefficient is related to the down-type Yukawa couplings.} in \fig{mnrad}

\section{Supersymmetry with Spontaneously Broken R--Parity}

\begin{figure}[t]
\centerline{\protect\hbox{\psfig{file=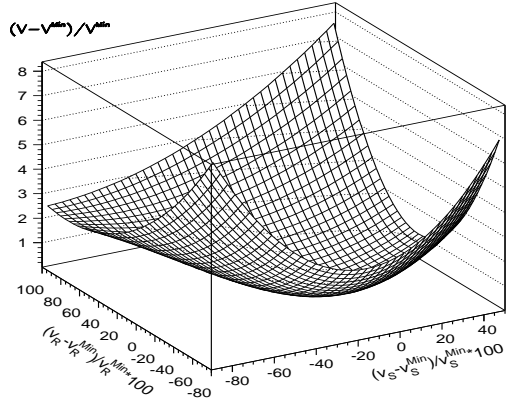,height=6cm,width=7.5cm}}}
\vglue -0.5cm
\caption{Scalar potential with radiative R--parity violation.}
\label{v1}
\end{figure}

A more satisfactory picture to R--parity violation would be one in
which it is conserved at the Lagrangian level but breaks spontaneously
through a sneutrino VEV \cite{beyond}.  Keeping the minimal \21 gauge
structure this also implies the spontaneous breaking of lepton number,
which is a continuous ungauged symmetry, and therefore the existence
of an associated Goldstone boson (majoron).  The breaking of R-parity
should be driven by {\sl isosinglet} right-handed sneutrino vacuum
expectation values (VEVS) \cite{MASIpot3} so as to avoid conflicts
with LEP observations of the invisible $Z$ width (in this case the
majoron is mostly singlet, and does not couple appreciably to the Z).
The theoretical viability of this scenario has been demonstrated both
with tree-level breaking of the electroweak symmetry and R--parity
\cite{MASIpot3}, as well as in the most attractive radiative breaking
approach \cite{RIV}. This is illustrated in \fig{v1}, taken from 
ref. \cite{RIV}.
The existence of the majoron, denoted by $J$, implies a novel Higgs
boson decay mode $H \to JJ$ which is a characteristic feature of \21
models with spontaneously broken R--parity, as well as in any model
with continuous global symmetries spontaneously broken at the weak
scale. This has an important impact on Higgs boson search strategies
at accelerators such as the Tevatron, LEP II, NLC and LHC
\cite{ebolep}.

Again in these models neutrinos will have mass, which will depend in
the details of the model. One possibility is to add only the $\nu^c$
right-handed neutrino superfields and give them masses {\sl a la
see-saw} \cite{sbrpothers}
\footnote{Strictly speaking, the addition of just one superfield suffices.
We prefer, however, to add them sequentially.}. It is conceptually
simpler, however, to give masses to the gauge singlets {\sl a la Dirac}
by adding two \21 singlet superfields $\nu^c$ and $S$ sequentially in
each generation \cite{MASIpot3,RIV,beyond}. Due to the original lepton 
number symmetry at the Lagrangean level \cite{SST} 
neutrinos are massless before breaking R--parity. The magnitude
of R--parity violating effects will be directly correlated with the
\nt mass which arises due to mixing with neutralinos, as mentioned
above. In this approximation the \ne and \nm remain massless. The
\nm may now get mass even in the tree-level approximation by mixing
with the singlets \cite{Romao92} or via SUSY loops.

Another class of models with spontaneous breaking of R--parity
consists of models where the gauge symmetry contains lepton number,
such as heterotic string inspired $E_6$ models with Calabi-Yau
compactifications or left-right symmetric models. These have been
discussed in the literature, see ref. \cite{RPCHI,RPLR}. The most
important difference with the \21 models is that in this case there is
no physical Goldstone boson (majoron) associated to the breaking of
lepton number since it is absorbed by the Higgs mechanism as the
longitudinal mode of some extra $Z^\prime$ gauge boson. 

In the following few sections I will illustrate with some examples the
potential of the present and future colliders in testing supersymmetry
with spontaneous or bi-linear breaking of R--parity under the
assumption that neutrinos acquire mass only due to R--parity
violation.  For simplicity we will refer to these models generically
as RPSUSY models. The characteristic feature of these models is that
the pattern of R--parity breaking interactions is determined in terms
of relatively few new parameters in addition to those of the MSSM (one
in the simplest reference model \cite{epsrad,DJV}).  This allows for a
systematic discussion of the potential of new colliders in searching
for broken R--parity SUSY signals. Many of the implications already
appear at the level of the truncated version of the model in which the
bi-linear term mimics the violation of R--parity in an effective
sense.  As mentioned previously the bi-linear model is consistent in
its own right, at least for a given range of \mnt values, say between
100 keV to an MeV
\footnote{Lighter \nt will not decay efficiently in order to cope 
with the cosmological critical density limits, while heavier ones may
have problems with primordial nucleosynthesis.}  

\begin{figure}[t]
\centerline{\protect\hbox{\psfig{file=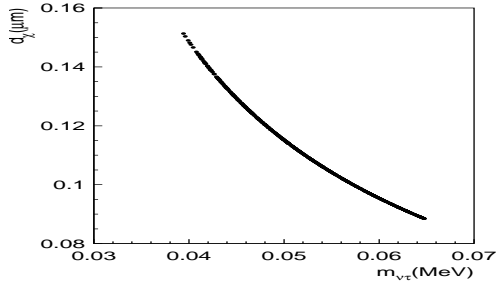,height=3.8cm,width=7cm}}}
\vglue -0.5cm
\caption{Typical neutralino decay path.}
\label{life}
\end{figure}

Before we start the phenomenological discussion, we note that, even
with relatively small strength of R--parity breaking interactions the
lightest neutralino is expected to decay inside the existing particle
detectors, for the typical energies of interest. This is illustrated
in \fig{life}, taken from ref. \cite{RIV}.
Although it refers to a particular model with spontaneous radiative
breaking of R--parity, similar features arise also in the bi-linear
model, or models with spontaneous breaking of R--parity in which
lepton-number is part of the gauge symmetry. In this case the
neutralino is likely to be the LSP and will decay with a sizeable
branching ratio into visible channels.  As a result the corresponding
effects can be quite striking experimentally, to the extent that the
missing momentum signature of the LSP in the MSSM will be
substantially diluted in favour of the appearance of novel exotic
signatures typically characterized by high fermion multiplicities (see
below).

\section{R--Parity Violation at LEP}

The requirement that SUSY is broken effectively at the weak scale
implies that SUSY particles are expected to exist at this scale, thus
it makes sense to search for SUSY signatures at colliders such as the
present Tevatron and LEP II, as well as the future LHC and NLC
colliders.

In the MSSM the usual neutralino pair-production process,
\begin{equation}
\label{chichi}
e^+ e^- \to  \chi \chi
\end{equation}
where $\chi$ denotes the lightest neutralino, leads to no
experimentally detectable signature (other than the contribution to
the Z invisible width if $m_Z > 2 m_\chi$), as $\chi$ escapes the
apparatus without leaving any tracks. The simplest process that leads
to a zen-event topology, with particles in one hemisphere and nothing
on the opposite, requires the production of $\chi$ associated to
$\chi^\prime$, the next-to-lightest neutralino, i.e. $ e^+ e^-
\to \chi \chi^\prime$.

In broken R--parity models the $\chi$ may decay into charged
particles, so that \eq{chichi} can lead to zen-events in which one
neutralino decays visibly (leptons and jets) and the other
invisibly. The topology is the same as in the MSSM but the
corresponding rates can be larger than in the MSSM and may occur below
the threshold for $\chi^\prime$ production.  The missing momentum in
these models is carried by the \nt or by majorons.  Another
possibility for zen events in RPSUSY is the process $ e^+ e^- \to \chi
\nu_\tau$. Since this violates R--parity, the rates are 
somewhat smaller, but might be observable at LEP I.

\begin{figure}[t]
\centerline{\protect\hbox{\psfig{file=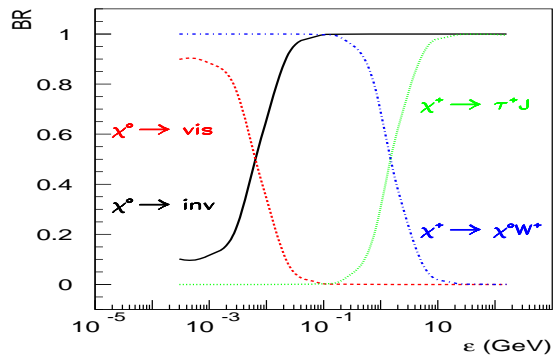,height=5cm,width=8.5cm}}}
\vglue -0.3cm
\caption{Typical neutralino and chargino decay branching ratios as a
function of $\epsilon_3 $.}
\label{br}
\end{figure}

For the sake of illustration we exhibit in \fig{br} typical values of
the branching ratios of neutralinos and charginos, as a function of
$\epsilon$ for $\mu=150$ GeV, $M_2=100$ GeV, and $\tan\beta=35$.  For
neutralinos we exhibit its total visible and invisible branching
ratios, where we included in the invisible width the contributions
coming from the neutrino plus majoron channel ($\chi \to \nu$J), as
well as $\chi \to 3 \nu$.

\begin{figure}[t]
\centerline{\protect\hbox{\psfig{file=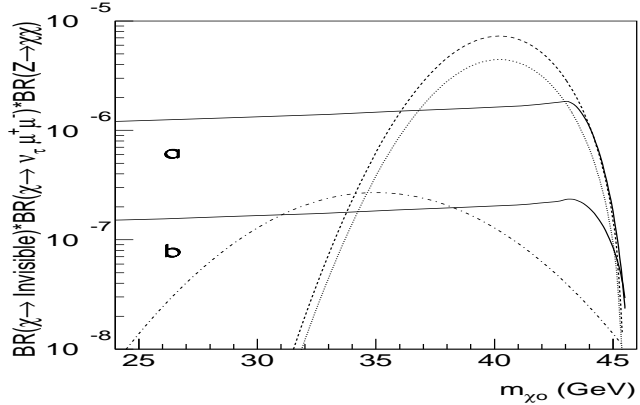,height=5.5cm,width=8.5cm}}}
\vglue -0.4cm
\caption{Limits on $BR (Z \to \chi \chi) BR(\chi \to \mu^+\mu^- \nu)$ 
versus theoretical expectations.}
\label{chichiab}
\vglue -.3cm
\end{figure}

In \fig{chichiab} we illustrate the sensitivity of LEP experiments to
leptonic signals associated to neutralino pair-production at the Z
peak in RPSUSY models. The signal topology used was missing
transverse momentum plus acoplanar muon events ($ p\!\!\!/_T +\mu^+
\mu^- $) arising from $\chi \chi$ production followed by $\chi$
decays.  The solid line (a) in \fig{chichiab} is the region of
sensitivity of LEP I data of ref. \cite{aleph95} corresponding to an
integrated luminosity of 82 $pb^{-1}$, while (b) corresponds to the
improvement expected from including the $e^+e^-\nu$ channel, as well
as the combined statistics of the four LEP experiments.  The dashed
line corresponds to the bi-linear model of explicit R--parity
violation, allowing \mnt values as large as the present limit, the
dotted one does implement the restriction on \mnt suggested by
nucleosynthesis \cite{bbnutau} and the dash-dotted one is calculated
in the model with spontaneous breaking of R--parity (majoron model).
The inclusion of semi-leptonic decays and of the updated integrated
luminosity already achieved at LEP would substantially improve the
statistics and thus the sensitivity to RPSUSY parameters.
The usual chargino pair-production process,
\begin{equation}
e^+ e^- \to \chi^+ \chi^-
\end{equation}
may also provide novel signatures which would not be possible in the
MSSM, as the neutralinos produced from chargino decays may themselves
decay into jets or leptons leading to exotic channels. 

Moreover, in \21 models with spontaneous violation of R--parity the
presence of the majoron implies the existence of two--body chargino
decays \cite{ROMA}
\begin{equation}
\label{tj}
\chi^\pm \to \tau^\pm + J
\end{equation}
In ref. \cite{tauj} chargino pair production at LEP II has been
studied in supersymmetric models with spontaneously broken
$R$--parity. Through detailed signal and background analyses, it was
shown that a large region of the parameter space of these models can
be probed.  The limits on the chargino mass depend on the magnitude of
the effective $R$--parity violation parameter $\epsilon$.  
\begin{figure}[t]
\centerline{\protect\hbox{\psfig{file=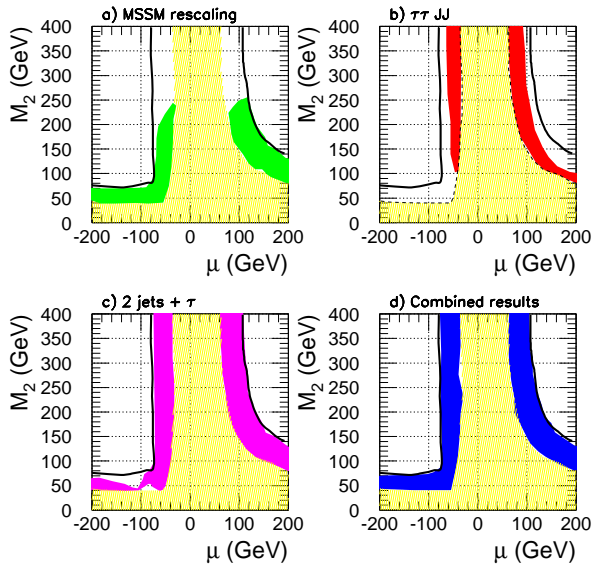,height=9cm,width=9cm}}}
\vglue -0.4cm
\caption{95\% CL excluded region in RPSUSY models in various analyses
(dark areas), and the combined excluded region for
$\protect\sqrt{s}=172$ GeV, and  300 pb$^{-1}$ integrated luminosity.}
\label{zone2-1}
\vglue -.3cm
\end{figure}
As $\epsilon \to 0$ we recover the usual MSSM chargino mass limits,
however, for $\epsilon$ sufficiently large, the bounds on the chargino
mass can be about 15 GeV weaker than in the MSSM due to the dominance
of the two-body chargino decay mode \eq{tj}. This happens because
there is an irreducible background from W-pair production with each $W
\to \tau \nu$. 

Although the \nt can be quite relatively heavy in these models, it is
consistent with the cosmology critical density \cite{KT} as well as
primordial nucleosynthesis \cite{unstable,DPRV}, due to the existence
of the majoron which opens new \nt decay and annihilation channels
\cite{fae}. The small mass difference between \ne and \nm may lead
to an explanation of solar neutrino deficit by resonant \ne to
\nm conversions \cite{MSW}. In this model one may regard the the 
R--parity violating processes as a tool to probe the physics
underlying the solar neutrino conversions \cite{Romao92}. For example, the
rates for some RPSUSY rare decays can be used in order to discriminate
between large and small mixing angle MSW solutions to the solar neutrino
problem \cite{MSW}.

\section{R--Parity Violation at LHC}

\begin{figure}[t]
\centerline{\protect\hbox{\psfig{file=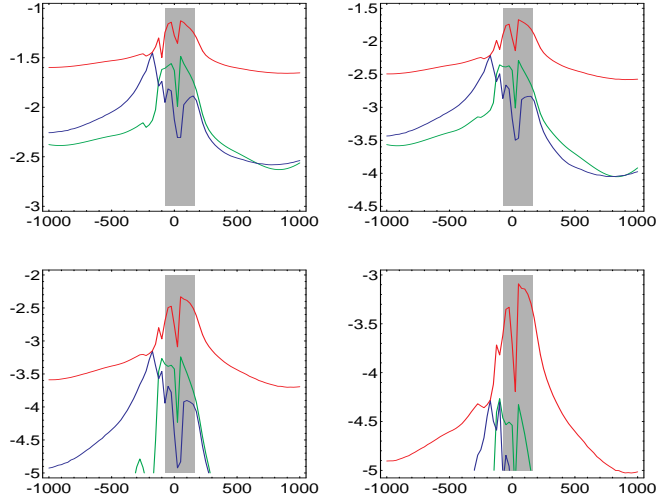,height=9cm,width=9cm}}}
\vglue -0.5cm
\caption{Multi-lepton rates at LHC in various RPSUSY models. }
\vglue -.4cm
\label{Clepmlt02}
\end{figure}

It is also possible to find manifestations of R--parity violation at
the super-high energies available at hadron super-colliders such as
the Tevatron and the LHC. If SUSY particles, gluinos and squarks, are
pair produced at hadron collisions, their subsequent cascade decays
will not terminate at the lightest neutralino but it will further
decay.  To the extent that this decay is into charged leptons it will
give rise to a quite rich pattern of high multiplicity lepton
events. Such pattern of gluino cascade decays in RPSUSY models was
studied in detail in ref. \cite{gluino}. The conclusion is that
multi-lepton and same-sign dilepton signal rates which can be
substantially higher than those predicted in the MSSM. This is
illustrated in \fig{Clepmlt02}, which shows the branching ratios for
various multi-lepton signals (summed over electrons and muons) with
the 3-, 4-, 5- and 6-leptons, for $\tan \beta = 2$, with other
parameters chosen in a suitable way (see ref. \cite{gluino} for
details).  We show a) the 3-lepton, b) the 4-lepton, c) the 5-lepton
and d) the 6-lepton signal for the MSSM (full line), the majoron-model
(dashed line) and the bi-linear model (dashed-dotted line). The shaded
area will be covered by LEP II. Note, for example, that for $\mu < 0$
the 5-lepton signal is much larger in the majoron-model than in the
MSSM, giving about 30 to 1200 events per year for an LHC luminosity of
$10^{5}pb^{-1}$.  The 6-lepton signal has a rate up to $5 \times
10^{-5}$ in the range $-300$~GeV$< \mu < -80$~GeV giving 125 events
per year. The multi-lepton rates would be even higher in the bi-linear
model.
Although with smaller rates, one also expects in RPSUSY models the
single production of the SUSY states in hadron collisions. For example
in ref. \cite{RPLHC} the single production of weakly interacting SUSY
fermions (charginos and neutralinos) via the Drell-Yan mechanism was
studied. 

\section*{Acknowledgements}

Supported by DGICYT grants PB95-1077, HP-1997-0039 and HU-1997-0046
and by EEC under the TMR contract ERBFMRX-CT96-0090.

\bibliographystyle{ansrt}

\end{document}